\documentclass[
 aps, pra,
 amsmath,amssymb,
 12pt,
 final,
tightenlines,
 nofootinbib,
 superscriptaddress,
 ]
{revtex4}

\usepackage[utf8x]{inputenc}
\UseRawInputEncoding
\usepackage[english]{babel}
\usepackage{graphicx}

\begin{document}

\title{High resolution  optical spectra of the dormant LBV star P\,Cyg}

\author{V.G.~Klochkova, V.E.~Panchuk and  N.S.~Tavolzhanskaya}
\affiliation{Special  Astrophysical Observatory,  Nizhnij Arkhyz, 369167 Russia}

\sloppypar
\vspace{2mm}
\noindent

\begin{abstract} 
High resolution optical spectra (R = 60 000) of the  LBV star P\,Cyg  beyond   outburst
were obtained on the 6-meter BTA telescope in the wavelength range 477--780\,nm. We
perform a detailed identification of different types lines (photospheric absorptions, 
permitted and forbidden emissions, components of lines with  P\,Cyg type profiles), 
and studied the variability of their profiles and radial velocities.
The average radial velocity from positions of forbidden emissions ([N\,II]\,5754.64, 
[FeII]\,5261.62, [FeII]\,7155.14 and [NiII]\,7377.83\,\r{A}) is accepted as the system
velocity Vsys=$-34\pm1.4$\,km/s. About a dozen photospheric absorptions of CNO triad ions 
and Si\,III are found, their stable position, Vr(abs)=$-73.8$\,km/s, shifted relative 
to at $-40$\,km/s, indicates that these absorbtions are formed in the pseudophotosphere
region. The high-excitation emissions ([OI]\,5577, 6300, 6363\,\r{A}, [OIII] 4959 and
5007\,\r{A}, as well as HeII\,4686\,\r{A}) are absent in the spectra.
The radial velocity Vr(DIBs)=$-11.8$km/s according to the position of numerous DIBs
is consistent with the position of the interstellar components of the D-lines NaI and
KI forming in the galactic Perseus arm. A color  excess E(B-V)=0.34+/-0.03 mag and
interstellar absorption Av=1.09 mag were determined by measurements of equivalent
widths of nine DIBs. 
\end{abstract}

\maketitle

\section{Introduction}
This paper is dedicated to the detailed study of the optical spectrum of the blue 
supergiant P\,Cyg (Sp=B1–2\,Ia–0ep). The history of P\,Cyg’s brightness behavior
in the 17th--18th centuries, following a sudden increase in its brightness by 
3 magnitudes, recorded in  the early 1600s, is outlined in the extensive publication
by de~Groot~\cite{Groot}. The documented episode of active state over several years later 
led to the classification of P\,Cyg as the LBV (Luminous Blue Variable)
star  nearest to the solar system. The concept of ``Luminous Blue Variable'' and the 
abbreviature LBV were introduced by~\citet{Conti} in 1984. Photometric monitoring of
P\,Cyg over the next 400 years shows that the star mostly remains in a quiet 
(``dormant'')  state.

The detailed history of the photometric behavior of the first historically known 
and nearest  LBV star, P\,Cyg, as well as a summary of the  results of spectroscopy 
of this star, are presented by the authors in reviews~\cite{HD, Israelian}. Spectroscopy 
of P\,Cyg  began almost three centuries after its massive outburst in 1600. In 1895, Campbell, 
in his review~\cite{Campbell} dedicated to stars whose spectra contain both bright and dark hydrogen
lines, proposed selecting stars with intense emissions shifted to the short-wavelength 
region of absorptions  as a separate type. The Pulkovo astronomer A.~Belopol’sky~\cite{Belopolsky}, 
who obtained spectra of P\,Cyg in 1899 with a prism spectrograph covering the wavelength 
range between H$\gamma$ and  H$\beta$   and identified several details, did not
observe significant changes between spectra taken from September~21 to October~1, 1899. 
However, variability in the spectrum of P\,Cyg was noted in the early 20-th century~\cite{Merrill}. 
Later, Adams and Merrill~\cite{AM} found variations in intensity and structure of absorptions
in the star spectrum.

High mass loss rates and eruptive events, driven by the high luminosity-to-mass ratio 
of stars, lead to the formation of gas nebulae in the circumstellar environment of 
massive stars. As a result of these processes, the surface of such stars is heavy obscured, 
and their  observed features (primarily the character of profiles of specific spectral details 
and their variability) mostly reflect manifestations of unstable processes in their envelopes. 
The authors of the article~\cite{Jager} emphasize that for stars of this type, parameters are 
not related to the stellar photosphere, but pertain to the wind region. Often, the seemingly 
peculiar spectrum of these stars is complex, formed in the circumstellar environment, and
may not contain details of the stellar atmosphere. The spectrum of the known star  P\,Cyg 
falls into this category, and the SIMBAD  mentions the presence of a complex emission 
nebula surrounding the star as one of the fundamental pieces of information about star.

By studying a sample of P\,Cyg spectra obtained in 1942--1964 on various telescopes, 
de~Groot~\cite{Groot} gathered data on line profiles and radial velocities for major spectral 
features (H\,I, He\,I, C\,II, N\,II, Si\,II, S\,II, Fe\,II, Fe\,III, and other ions). 
Analyzing the variability of their short-wavelength absorption components, the author 
concluded that they form in the expanding circumstellar environment. In the same work, 
an estimate of a high mass loss rate of  $2\times 10^{-4}\mathcal{M}_{\odot}/$year was made,
and pulsation was indicated as the cause of the star’s brightness variability. In 
a~subsequent  study, authors~\cite{Najarro}, through modeling optical and IR spectroscopy data, 
determined a set of key parameters for P\,Cyg, calculated theoretical profiles
of H\,I and He\,I lines, and significantly reduced the mass loss rate to 
$3\times 10^{-5}\mathcal{M}_{\odot}/$year.

The primary characteristics of the optical spectrum of P\,Cyg were identified in the 
1980s based on photographic spectra: powerful emissions of H\,I, He\,I with structured 
absorption  components~[11]. A spectral atlas was created by \citet{Markova} based on
 photographic  spectra with a relatively low signal-to-noise ratio (S/N).
Widely cited is another spectral atlas~\cite{Stahl1993, Stahl1994}, based on echelle 
spectra covering a wide wavelength range, recorded using a CCD. The significance of 
this work lies in the detailed identification of the P\,Cyg spectrum within a broad 
wavelength range,  around 500\,nm.
Analysis of data at moderate resolution (R=12\,000) allowed the authors~\cite{Stahl1994} 
to  identify variability in emission intensity up to 30\% and radial velocity
(30$\div$50\,km/s) on a timescale of several months. It is important to note that 
these authors found no clear correlation in the behavior of spectral and photometric
features. They also did not observe the splitting of absorptions found earlier~\citet{Markova}. 
It should be noted that the atlas~\cite{Stahl1993} is based on data obtained by averaging 
spectra taken on different dates from 1989 to 1991 with spectrographs at various observatories 
to enhance S/N. For a star with a variable spectrum, the averaging procedure can lead 
to the loss of some information about the behavior of line profiles and radial velocities.

Currently, P\,Cyg is a full-fledged prototype of hot, highly luminous, unstable 
stars (LBV). Its location on the Hertzsprung--Russell diagram near the luminosity
limit in the instability strip of LBV stars is illustrative (see, for example, the 
informative Figure~3 in the review~\cite{Weis}). The main parameters of the star 
are presented in the~\cite{Turner} and compared with the parameters of related LBV stars 
and LBV candidates  in the article~\cite{Mahy}, Table 1). Among related LBV stars, 
P\,Cyg stands out for two reasons. Firstly, along with $\eta$\,Car, it belongs to 
the two stars in our Galaxy  for which significant brightness changes have been 
recorded in the past. Secondly, P\,Cyg has an optical spectrum saturated with intense  
variable emission-absorption  profiles of H\,I, He\,I, N\,II, S\,II, Fe\,II, and other ions, 
indicating  matter outflow  due to effective and  variable wind.
This feature of P\,Cyg profiles led to the recognition of the well-known  now
spectral phenomenon -- the P\,Cygni-type profile (or inverse P\,Cyg). The richness
of P\,Cyg spectrum with emissions of different origins requires detailed spectroscopy
for its study. It is worth noting that in the early work of de~Groot~\cite{Groot}, the terms
``P\,Cyg-type profile'' and ``stars of the P\,Cyg type'' were introduced into spectroscopy 
practice. Later, Lamers~\cite{Lamers} expanded the circle of objects with P\,Cyg type profile 
features and introduced the concept of the ``astrophysical phenomenon of P\,Cyg-type 
profiles''.

The combination of observed photometric and spectral  specificities of P\,Cyg is presented 
in the article~\cite{Gent} based  on the extensive observations we already mentioned 
as~\cite{Groot}.  An important result of the study~\cite{Gent} is the conclusion about 
the absence of any certain  period of parameters changes: long-term data of weak photometric 
variability indicate a possible period range from 0.5~days to 18~years.

To date, several works have been published based on spectra of P\,Cyg obtained at various 
times, and even fewer high-resolution spectral data are available. The significance of the 
work~\cite{Jager}, based on a sample of high-resolution spectra  (R=80\,000)  and high S/N ratio,  
obtained during the nights of May~28 to June~4, 1999, is evident. 
The authors of this work concluded that there are two periods of variability: one
matching the photometric period of 17.3~days, and a longer one of about 100~days. Clearly, 
observations over a week are insufficient to study spectral variability. As noted
by the authors~\cite{Jager}, the search for variability in the parameters of such a star is a 
``tricky task''.

The lack of spectral data of the required quality and volume has prompted us to 
initiate long-term spectroscopy of P\,Cyg to provide the search for variability
in the profiles of spectral details and the velocity field based on homogeneous 
high-quality  spectral data. This task will require multiple and possibly multiyear 
spectroscopy  with  high spectral resolution across a wide wavelength range. In this 
article, we present the results of the first stage of the work conducted to refine 
the identification of spectrum details, measure radial velocities, and identify 
spectral variability based on observations of P\,Cyg in 2021--2022. Section~2 of 
this paper briefly describes the  methods of observations and data analysis. 
Section~3 presents the obtained results,  while Sections~4 and 5 provide a discussion 
of our results and their comparison with  previously published works, along with the 
main conclusions.

\section{ECHELLE SPECTROSCOPY at BTA}

The spectra used in this study were obtained with the echelle spectrograph NES~\cite{NES},
permanently  installed at the Nasmyth focus of the 6-m BTA telescope. The observation 
dates for the star are presented  in Table~1. The spectrograph is equipped with a CCD
matrix with a number of elements 4608$\times$2048, and each element has a size of 
0.0135$\times$0.0135\,mm; the readout noise is 1.8e$^{-}$. Monitoring of P\,Cyg is
conducted in the wavelength range  $\Delta\lambda=470\div778$\,nm. To minimize flux 
losses at the entrance slit, the echelle spectrograph NES is equipped with a star 
image slicer.  With the use of the slicer, each spectral order is repeated three times.
The spectral resolution of NES is R=$\lambda/\Delta\lambda\ge$60\,000. 
In the spectra of P\,Cyg, the signal-to-noise ratio, S/N, varies by several orders 
of magnitude, from the continuum level to the peaks of the strongest emissions. 
In the continuum near 5000\,\AA{}, the ratio S/N=300$\div$360 for different observation
dates.

Extraction of one-dimensional spectra from two-dimensional echelle frames was performed
using a modified ECHELLE context in the ESO~MIDAS  package, accounting for the geometry 
of the echelle frame. All details of the procedure are described by authors~\cite{Echelle}.
Cosmic ray traces were removed using the standard method--by median averaging a pair of 
sequentially obtained spectra. Th-Ar lamp was used for wavelength calibration. All subsequent
steps in processing one-dimensional spectra were carried out using the current version of 
the DECH package~\cite{Dech}.

The systematic error of heliocentric radial velocity measurements based on a set 
of numerous telluric details and interstellar lines of the Na\,I doublet does not 
exceed 0.25\,km/s for a single line. For the averaged velocity values in Table~1, 
the errors of the mean depend on the type and number of measured lines.

Identification of features in the P\,Cyg spectra was performed using line lists 
from articles based on spectroscopy at BTA\,+\,NES of related hot, high-luminosity
stars, including stars with the B[e] phenomenon~\cite{CKMir, KloChen, Mirosh, Schulte12}. 
Additionally, information  from the VALD database (see~\cite{VALD} and references therein) 
was used for the   identification of several spectral details. 

\begin{table*}[ht!]
\medskip
\caption{Results of heliocentric velocity measurements in P\,Cyg spectra based 
    on measurements of a set of different line types}
\begin{tabular}{l| c| c|  c   }
\hline
                 & \multicolumn{3}{c}{Vr,  km/c} \\
\hline
Type of features & 26.10.2021 &  08.09.2022  &  09.09.2022  \\
\cline{1-4}
Pure absorbtions &$-74.4\pm 1.3$\,(13) &$-74.0\pm 2.3$\,(12)&$-73.0\pm 1.7$\,(13)\\ [-7pt]
Pure emis permit &$-38.3\pm 3.7$\,(23) &$-44.5\pm 1.8$\,(27)&$-43.5\pm 1.0$\,(30) \\ [-7pt]
Pure emis forb   &$-48.7\pm 5.5$\,(8)  &$-42.3\pm 3.3$\,(8) &$-41.7\pm 2.6$\,(7) \\ [-7pt]
P\,Cyg em HI HeI &$-17.5\pm 1.5$\,(9)  &$-12.2\pm 4.4$\,(8) &$-18.1\pm 3.3$\,(8)\\  [-7pt]
P\,Cyg ab HI HeI &$-147.8\pm 9.1$\,(10)&$-139.8\pm 14.3$\,(10)&$-147.2\pm 8.9$\,(10)\\ [-7pt]
P\,Cyg other em  &$-23.1\pm 1.3$ (23)  &$-23.5 \pm 1.3$\,(19)&$-23.5 \pm 1.2$\,(25)\\ [-7pt]
P\,Cyg other ab  &$-97.3\pm 2.4$\,(29) &$-99.0\pm 2.1$\,(29)&$-95.1 \pm 2.2$\,(29)\\ [-7pt]
DIBs             &$-12.1\pm 0.5$\,(20) &$-11.5\pm 0.5$ (31) &$-12.1\pm 0.4$\,(24) \\
\hline
\end{tabular}
\end{table*}

\section{RESULTS}

\subsection{Main Properties of the Spectrum} 

Figures~1, 2, and 3 with fragments of the P\,Cyg spectrum obtained in 2021
characterize the optical spectrum of P\,Cyg as a combination of diverse 
emissions. Primarily, these are lines of neutral hydrogen and helium with 
intense profiles of the classical P\,Cyg type. All figures in the text 
are based on P\,Cyg spectra obtained with the NES spectrograph. In Fig.\,4, 
the relative intensity of the H$\alpha$ emission is  I/Icont$\approx$14 in 
2021 and reaches  I/Icont$\ge$18 in the 2022 spectrum.

\begin{figure}[ht!]
 \includegraphics[angle=0,width=0.7\textwidth,bb=30 40 720 540,clip]{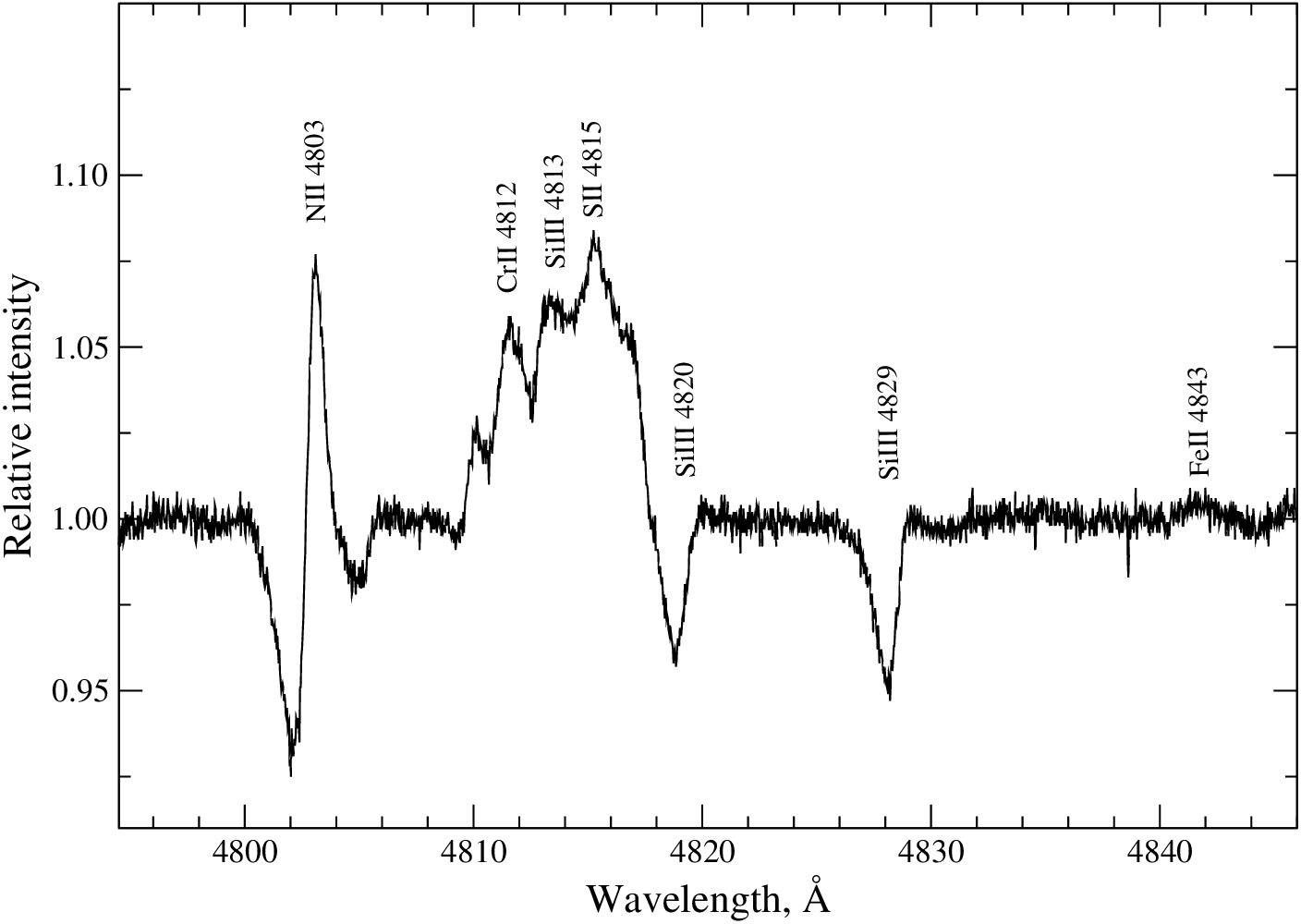} 
\caption{Fragment of the P\,Cyg spectrum}
\label{4800}
\end{figure}

\begin{figure}[ht!]
 \includegraphics[angle=0,width=0.7\textwidth,bb=30 40 720 540,clip]{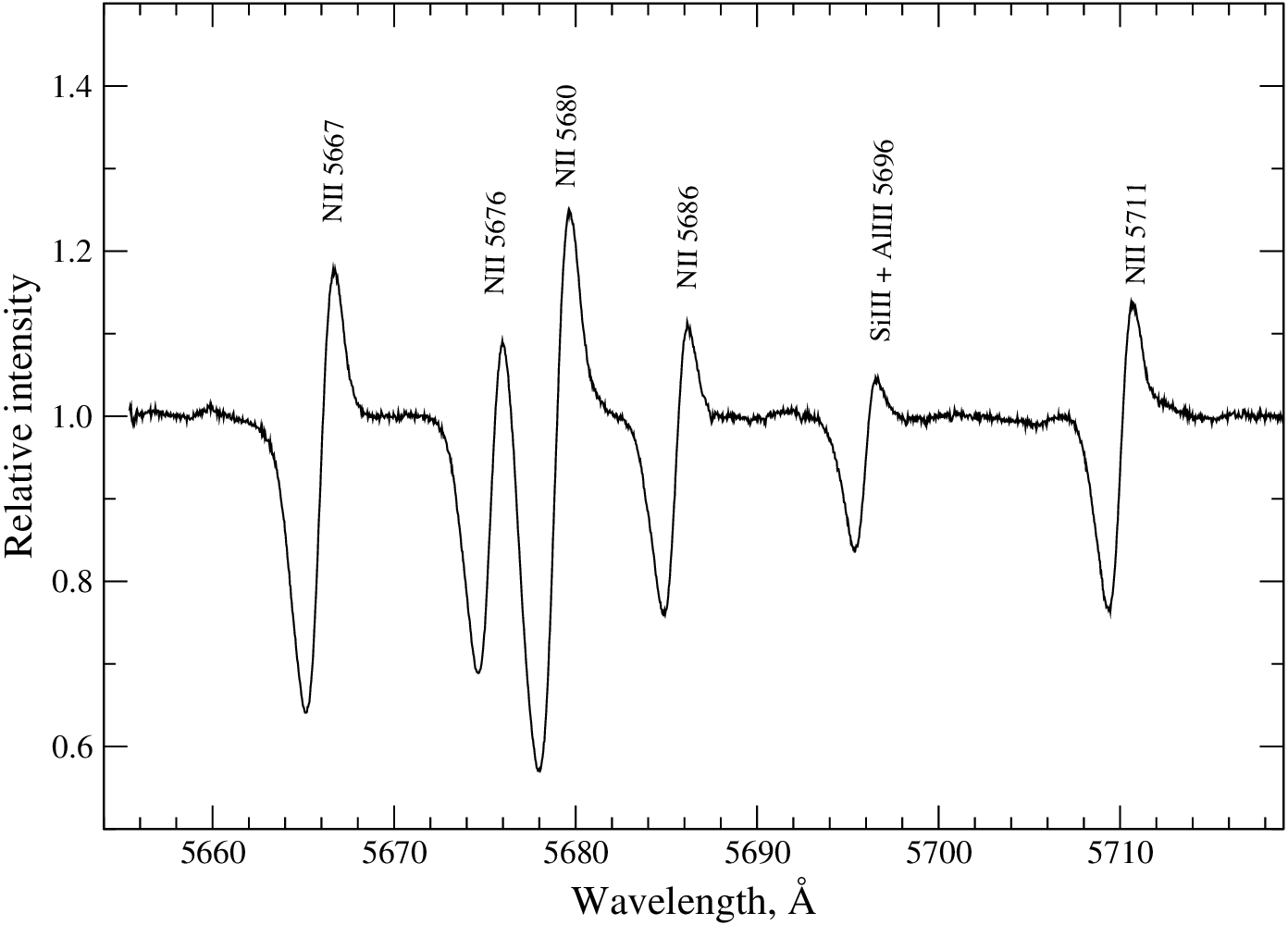}   
\caption{Fragment of the P\,Cyg spectrum with a set of intense N\,II lines with P\,Cyg-type profiles.}
\label{5680}
\end{figure}

\begin{figure}[ht!]
 \includegraphics[angle=0,width=0.7\textwidth,bb=30 40 720 540,clip]{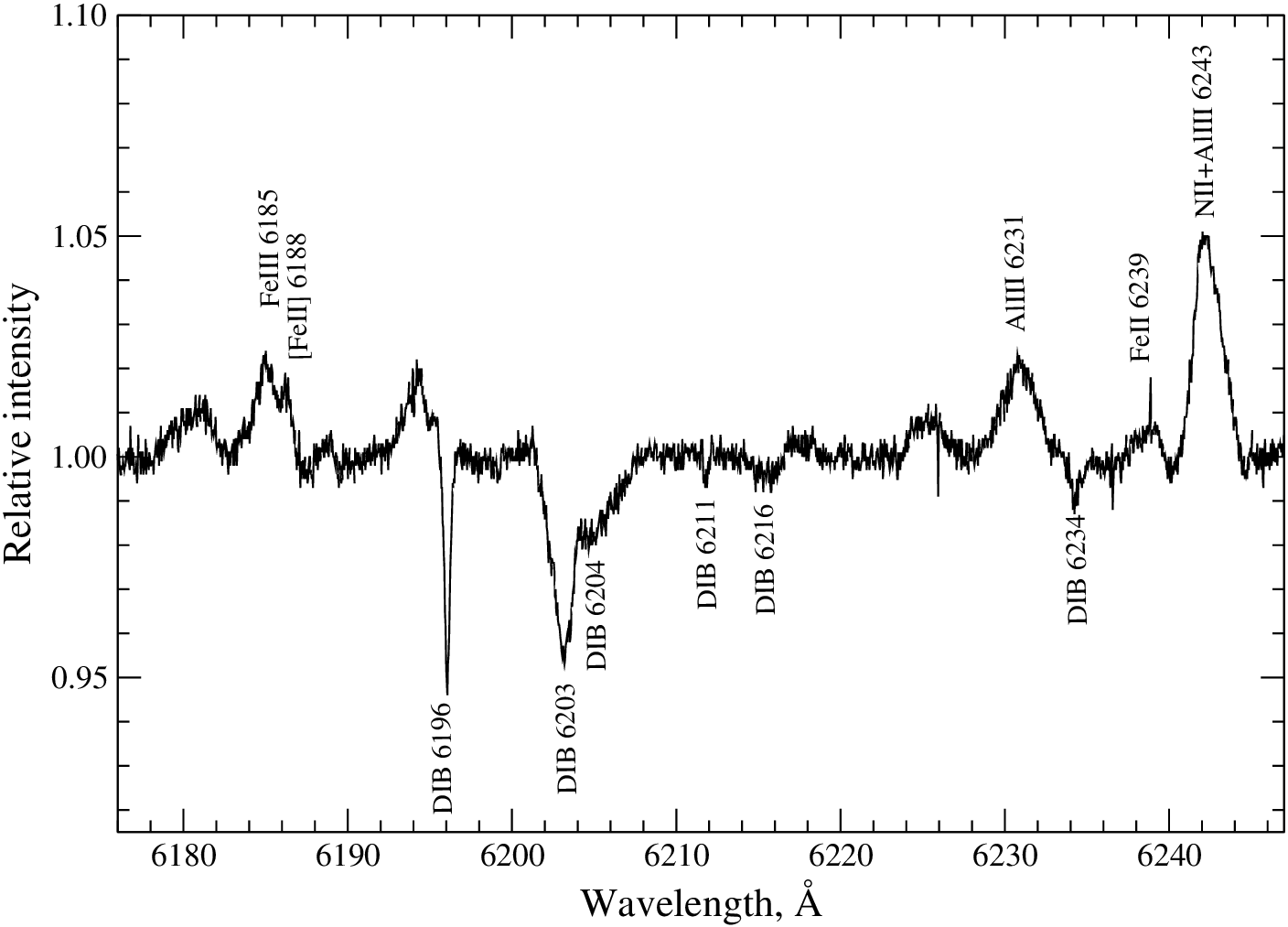} 
\caption{Fragment of the P\,Cyg spectrum containing both emissions and several DIBs.}
\label{6200}
\end{figure}

The  profiles of HI and HeI lines   in  spectra from different dates,
as shown in Figs.~4 and 5, indicate variability in the intensity of peak emission 
and absorption components. As indicated in Table~1, the positions of the emission 
components of HI and HeI differ and noticeably change over time. The differences
in the positions of absorption components  are much higher: the standard deviation 
exceeds 10\,km/s for the available dates.

\begin{figure}[ht!]
\includegraphics[angle=0,width=0.49\textwidth,bb=20 20 550 680,clip]{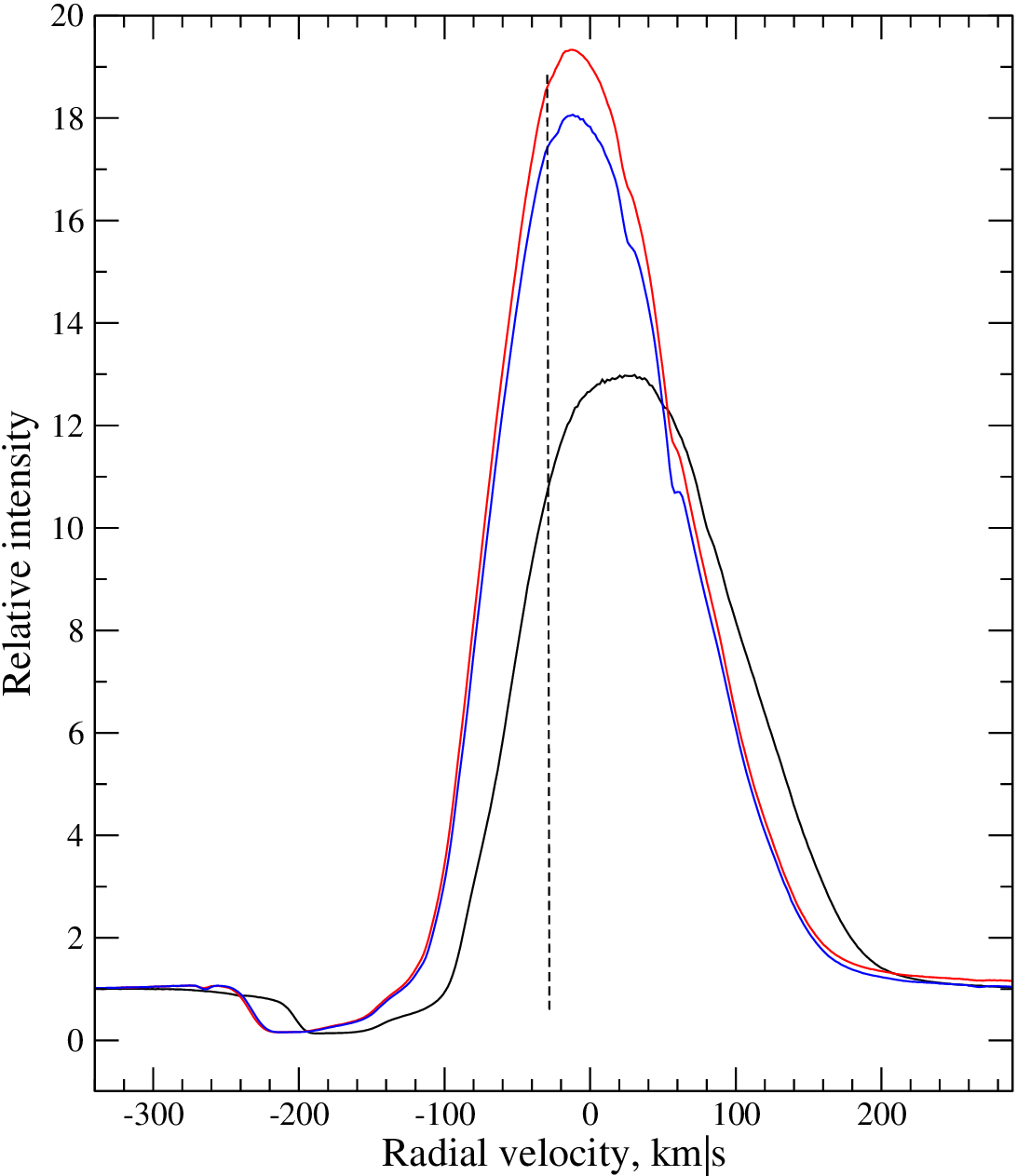}  
\includegraphics[angle=0,width=0.49\textwidth,bb=20 20 550 680,clip]{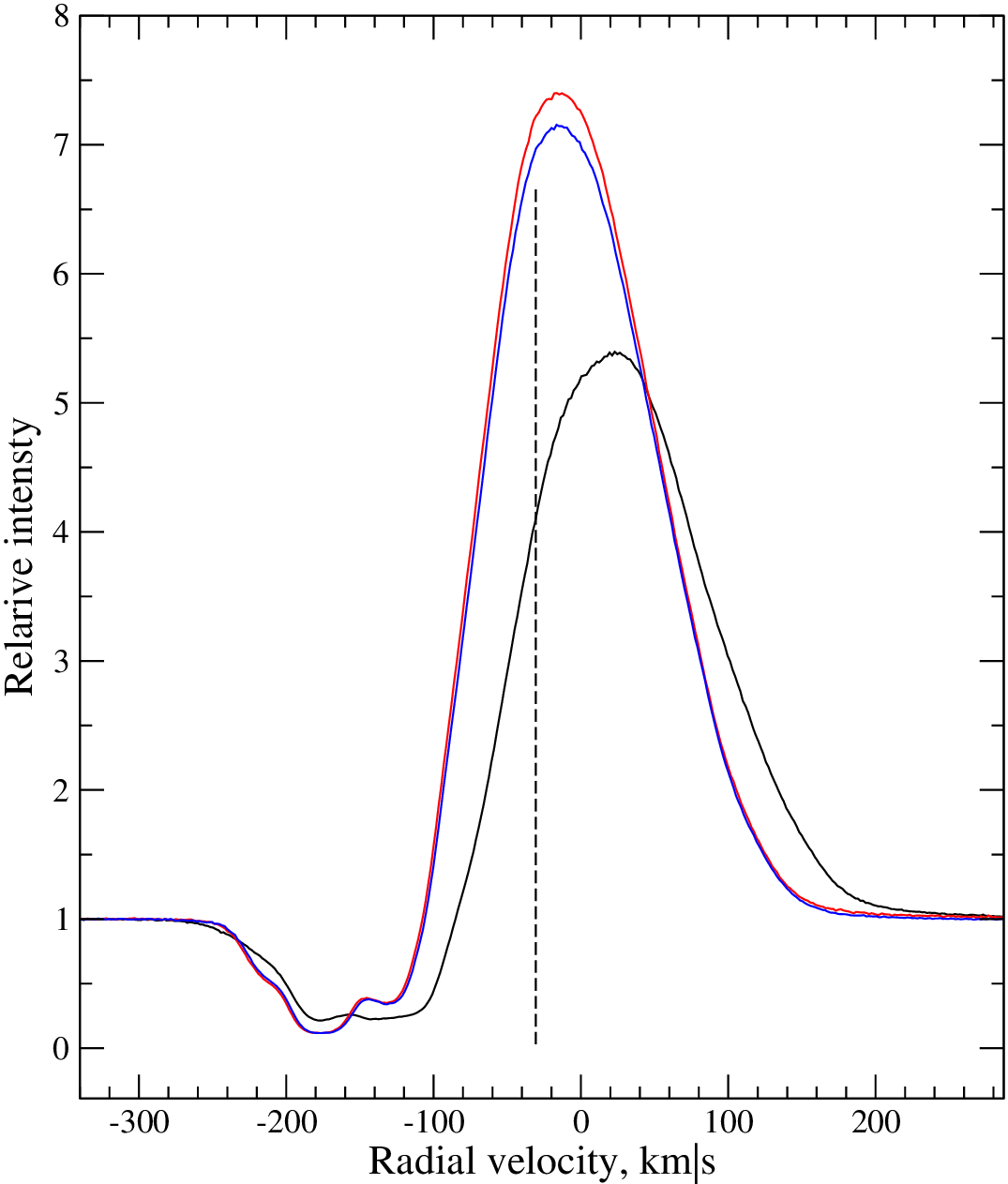}  
\caption{Variability of the H$\alpha$ (left panel) and H$\beta$ (right panel) profiles in the spectra 
of P\,Cyg on October~26, 2021 (black line)
and September~9, 2022 (red and blue lines, respectively). On this and subsequent figures with line 
profiles, the position of the dashed vertical line corresponds to the accepted value of the systemic 
velocity Vsys=$-34\pm1.4$\,km/s. Here and in subsequent figures with line profiles, the heliocentric 
radial velocity Vr in km/s is indicated along the abscissa axis.}
\label{Halpha}
\end{figure}

\begin{figure}[ht!]
\includegraphics[angle=0,width=0.45\textwidth,bb=10 20 550 680,clip]{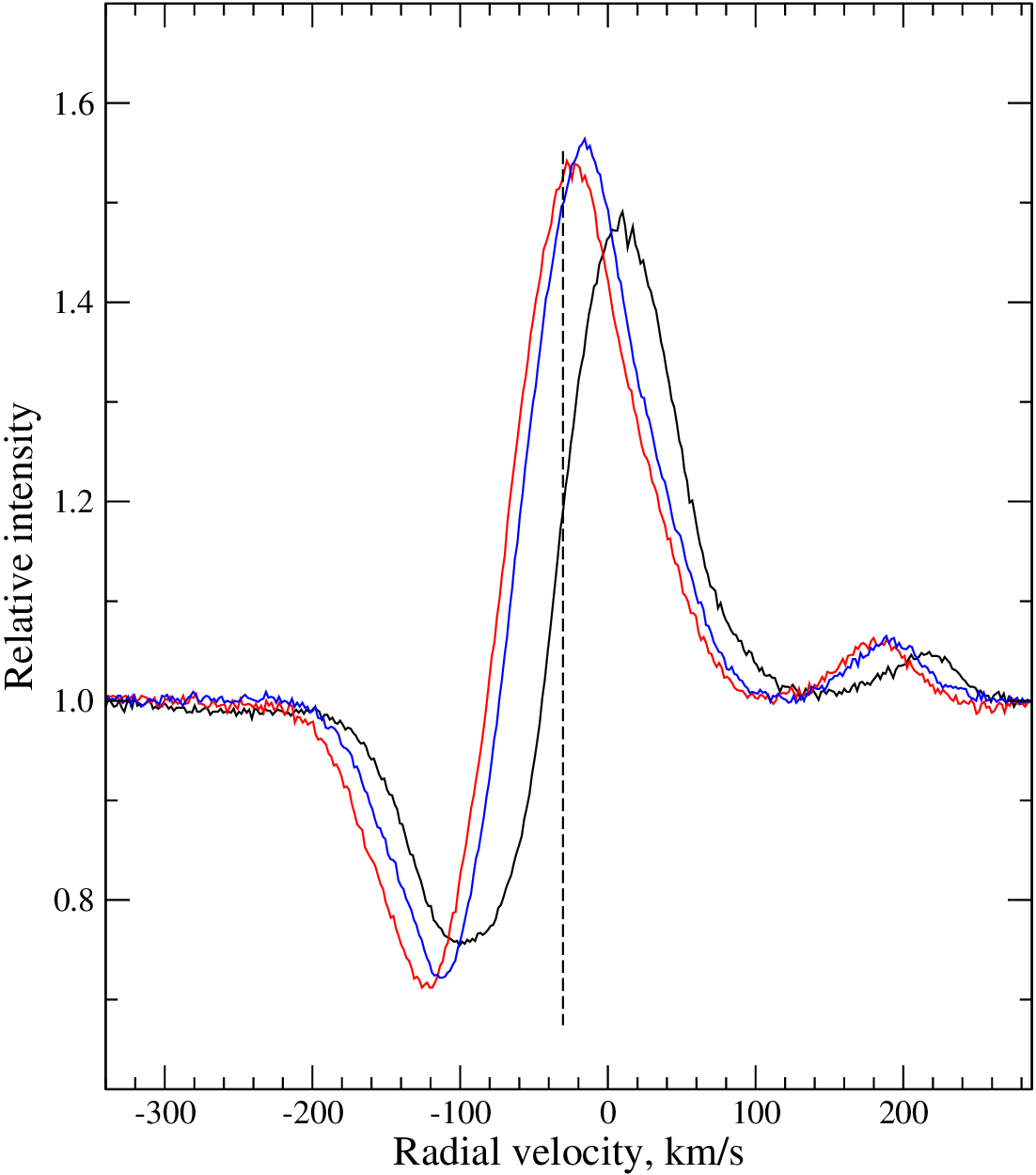}  
\includegraphics[angle=0,width=0.45\textwidth,bb=10 20 550 680,clip]{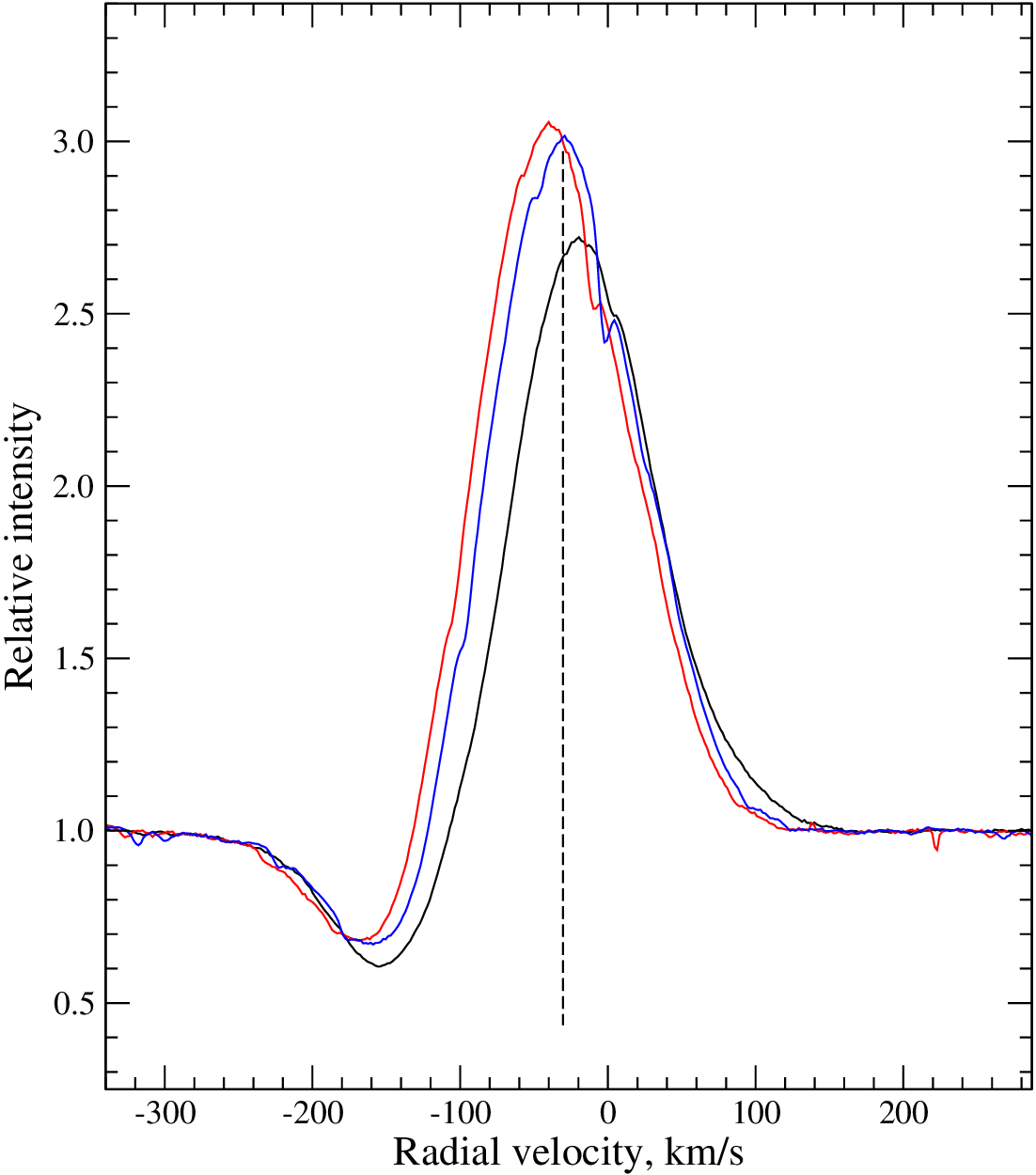}  
\caption{Profiles of two HeI lines in P\,Cyg spectra obtained in 2021 (black line) and 2022
(red and blue lines, respectively). Left: HeI~4713\,\AA{} line, right: HeI\,7065\,\AA{}  line.}
\label{HeI}
\end{figure}

It is evident that the dominant contribution to the high velocity dispersion of 
wind absorptions is not measurement errors, but the real variability of this 
velocity and the presence of structure  in these components. A quantitative study 
of this variability will be carried out in the future as the necessary
collection of star spectra accumulates. From the data in the penultimate row of 
Table~1,  the stability of the positions of emissions of the P\,Cyg type profiles 
for other elements (about three dozen emissions of NII, CII, Si\,II, Al\,III,
[FeII]) is evident. As shown in Figs.~2, 3, and 6, the peak emissions of this
series of lines also significantly exceed the continuum level. It is the radiation 
pressure caused by the plenty of such emissions in the Balmer and Lyman continuum 
that is the cause of the effective stellar wind according to~\cite{Lamers}.

The earliest studies on P\,Cyg spectroscopy noted the presence of separate wind 
components in the absorptions of H\,I and He\,I, caused by recurrent matter 
dumps. These so-called DACs (Diffuse Absorptional Components) were termed ``splitting'' 
in the early papers by Markova~\cite{Markova2000}. Markova noted the presence of this 
effect in the  lines of H\,I and He\,I and she expected splitting even in Fe\,III 
emissions. Interestingly, the atlas~\cite{Stahl1994} notes the complete absence of
this effect, presumably due to low spectral  resolution. Our observational data 
reliably record up to three separate wind components  in the radial velocity interval 
of Vr$\approx -(140\div250)$\,km/s only in the H$\alpha$  and H$\beta$ profiles presented 
in Fig.\,4. The splitting effect is especially clearly seen in the H$\beta$ profiles, where 
the intensity difference between the peak  values of emission and wind absorptions is 
many times smaller than in the H$\alpha$ profiles.

\begin{figure}[ht!]
\includegraphics[angle=0,width=0.6\textwidth,bb=20 70 550 680,clip]{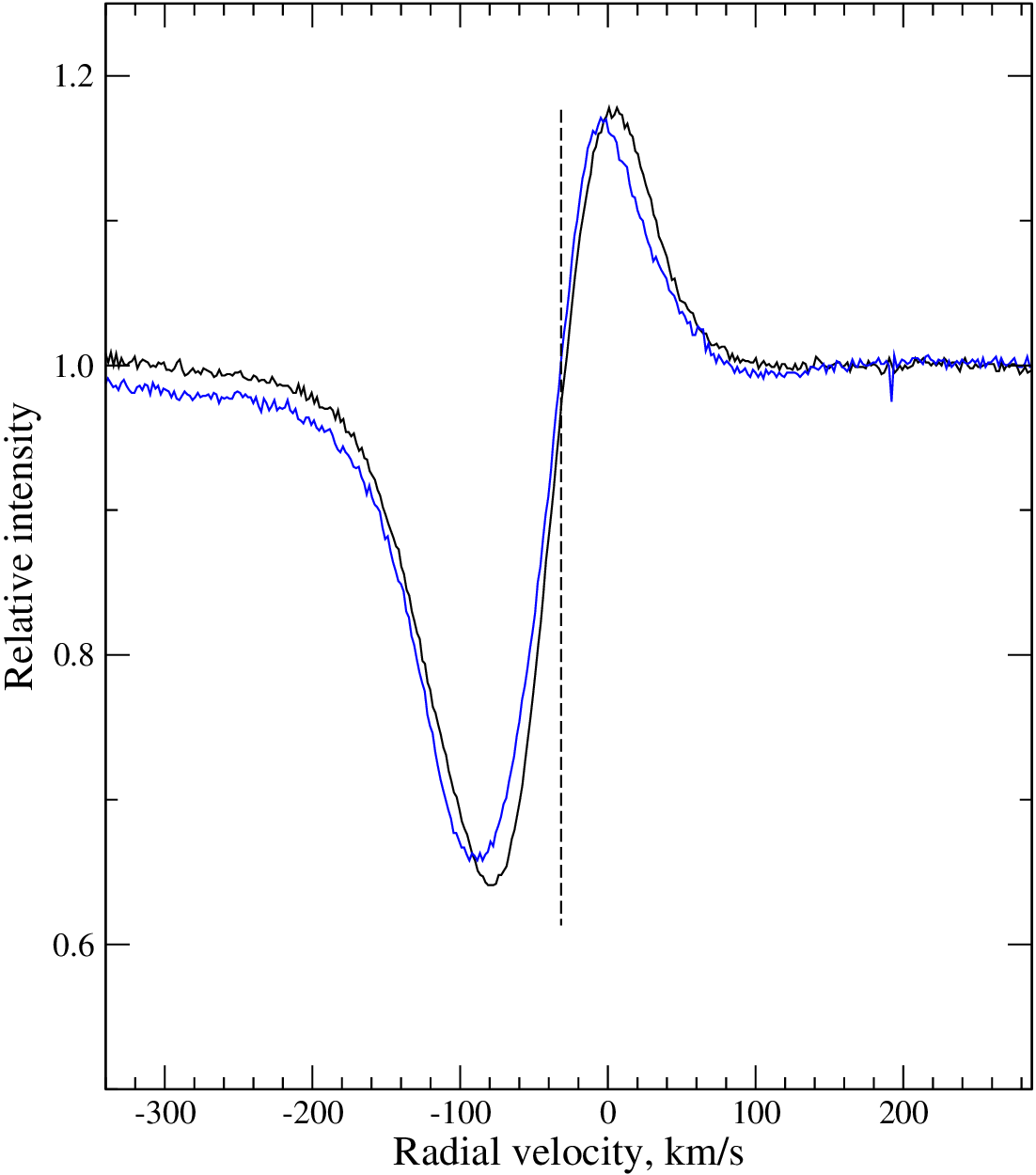}  
\caption{Variability of the profile of one of the most intense NII lines at 5666\,\AA{} in P\,Cyg
spectra obtained on October~26, 2021 (black line) and September~9, 2022 (blue line).}
\label{5666}
\end{figure}

We should emphasize that there is no splitting of profiles for lines of other elements 
in P\,Cyg spectra. This is illustrated, in particular, in Fig.\,2, showing a
fragment of the P\,Cyg spectrum containing a set of intense and unblended N\,II lines 
with P\,Cyg type profiles, as well as in Fig.\,6 with the variable profile of one
of these lines, NII\,5666\,\AA{}, in spectra from different dates of our observations.
The P\,Cyg spectrum also lacks  forbidden emissions for nebulae [OI]\,5577, 6300, 
6363\,\AA{}, [OIII]\,4959 and 5007\,\AA{}. There is also no emission of HeII~4686\,\AA{}, 
the absence of which in the P\,Cyg spectrum was emphasized by O. Struve~\cite{Struve}, 
who identified numerous lines of NII, OII, Si\,III, SII, FeII in the optical range.
In our spectra, we did not see forbidden emissions [CaII]~7291 and 7324\,\AA{}, which 
in the spectra of selected high-luminosity stars indicate the presence of a circumstellar 
disk. Spectra of the extremely luminosity star MWC\,314~\cite{Mir1998} or yellow hypergiant 
V1302\,Aql~\cite{KloChen} are excellent examples. 

Powerful emissions of H\,I and He\,I in the P\,Cyg spectrum are combined with numerous 
weak emissions of NII, CII, Si\,II, Si\,III, FeII, FeIII, etc., having a wind
absorption component. These permitted emissions coincide by 30\% with the lines from 
the list in the paper~\cite{Markova1997}. Figure~2 well represents the features of 
this line sample:  the intensity of these envelope emissions rarely exceeds 15--20\% 
above the local  continuum level. Average velocities are given in Table~1 for the 
positions of emission  (P\,Cyg~other~em) and absorption (P\,Cyg~other~abs) components 
of these lines. The absorption component of these lines is shifted by less than $-100$\,km/s. 
The features  of this type lines are well illustrated in Fig.\,6, which shows the profile of
a typical and sufficiently intense NII\,5666\,\AA{} line.

\begin{figure}[ht!]
\includegraphics[angle=0,width=0.6\textwidth,bb=20 80 550 680,clip]{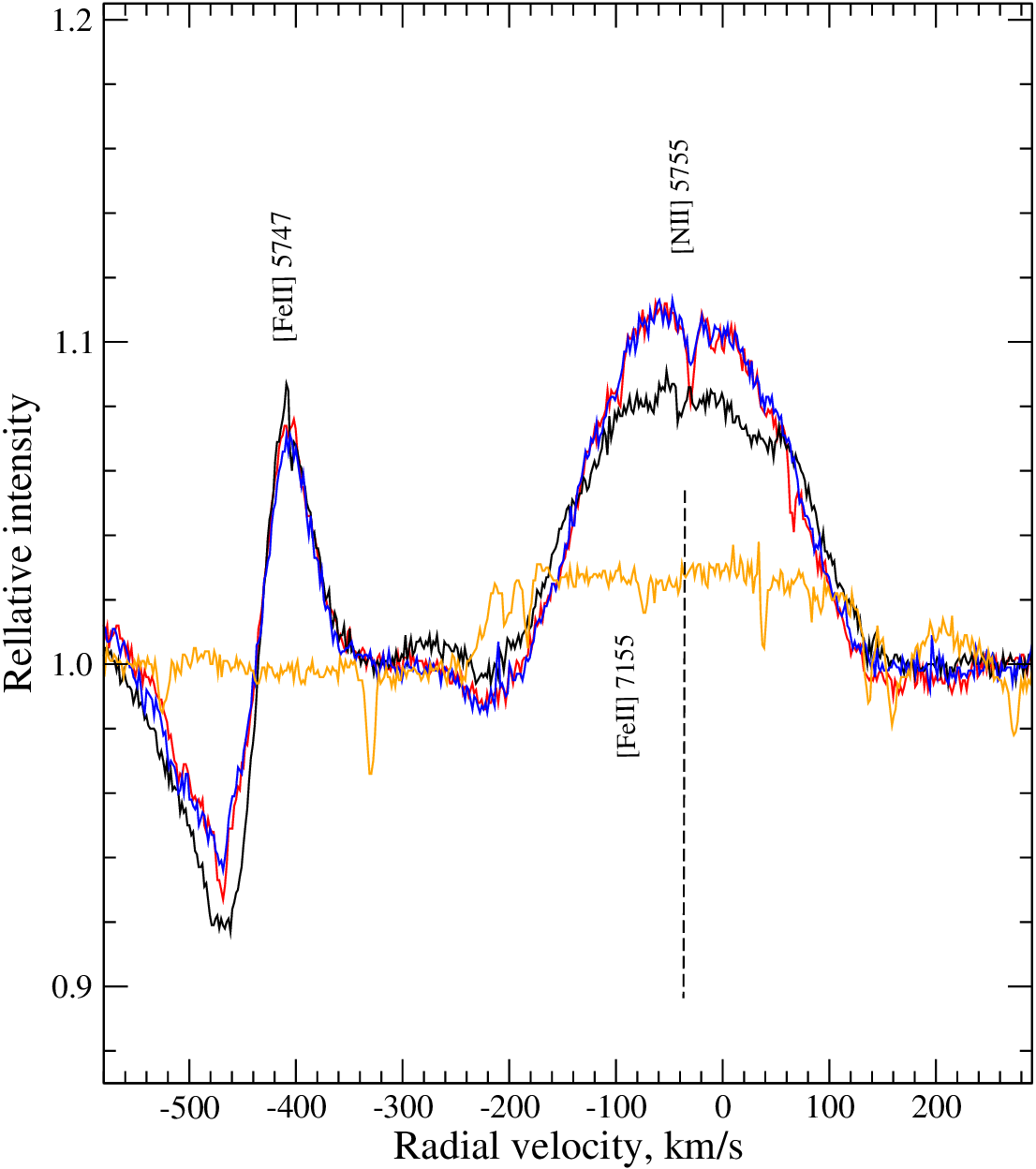}  
\caption{Profile of the forbidden line [NII]\,5755\,\AA{} in P\,Cyg spectra obtained in 2021~(black line) 
 and 2022~(red and blue lines).
The figure also includes the variable forbidden emission [FeII]\,5746.96\,\AA{}. The orange line 
represents the profile of a weak and broad [FeII]\,7155\,\AA{} emission.}
\label{5754}
\end{figure}

The intensity of forbidden emissions is even lower, the most noticeable of which 
is the forbidden nitrogen line [NII]\,5755\,\AA{} (see Fig.~7). This emission is
a crucial marker, the presence of which in the spectrum allows a priori assigning 
a high-luminosity star to the small family of LBV stars. The forbidden emission
[NII]~6583\,\AA{}, close in origin to the emission [NII]~5755\,\AA{}, is absent
in the P\,Cyg spectrum or blended with the CII~6582.9\,\AA{} emission.

\subsection{Radial Velocity Pattern}

The high  S/N$\ge 300$  ratio in our P\,Cyg spectra allowed to identify 
additional details compared to the photographic spectrum in the atlas~\cite{MZ}. 
A complicating factor in detailed identification of the emission spectrum 
is the low intensity of permitted emissions of metal ions, except for NII lines.
The nebula in the P\,Cyg system belongs to a rarely observed peculiar type: 
the authors of~\cite{Nota} indicated  that this faint spherical~\cite{Weis} nebula 
consists  of individual clumps with a size of 2--3~arcsec, distributed along an enevelope  
with a diameter of about 20~arcsec. This nebula is also characterized by an 
extremely small mass of ionized gas, 0.00092$\mathcal{M}_{\odot}$~\cite{Barlow}.
The small  volume of the gaseous envelope manifests itself in a combination of weak 
permitted and forbidden emissions of various origin. The emission profiles presented 
in Figs.~4, 5, and 8 demonstrate significant differences in the peak intensities of  
emission and absorption components while coinciding in the range of their Vr values. 

\begin{figure}[ht!]
\includegraphics[angle=0,width=0.48\textwidth,bb=10 20 550 680,clip]{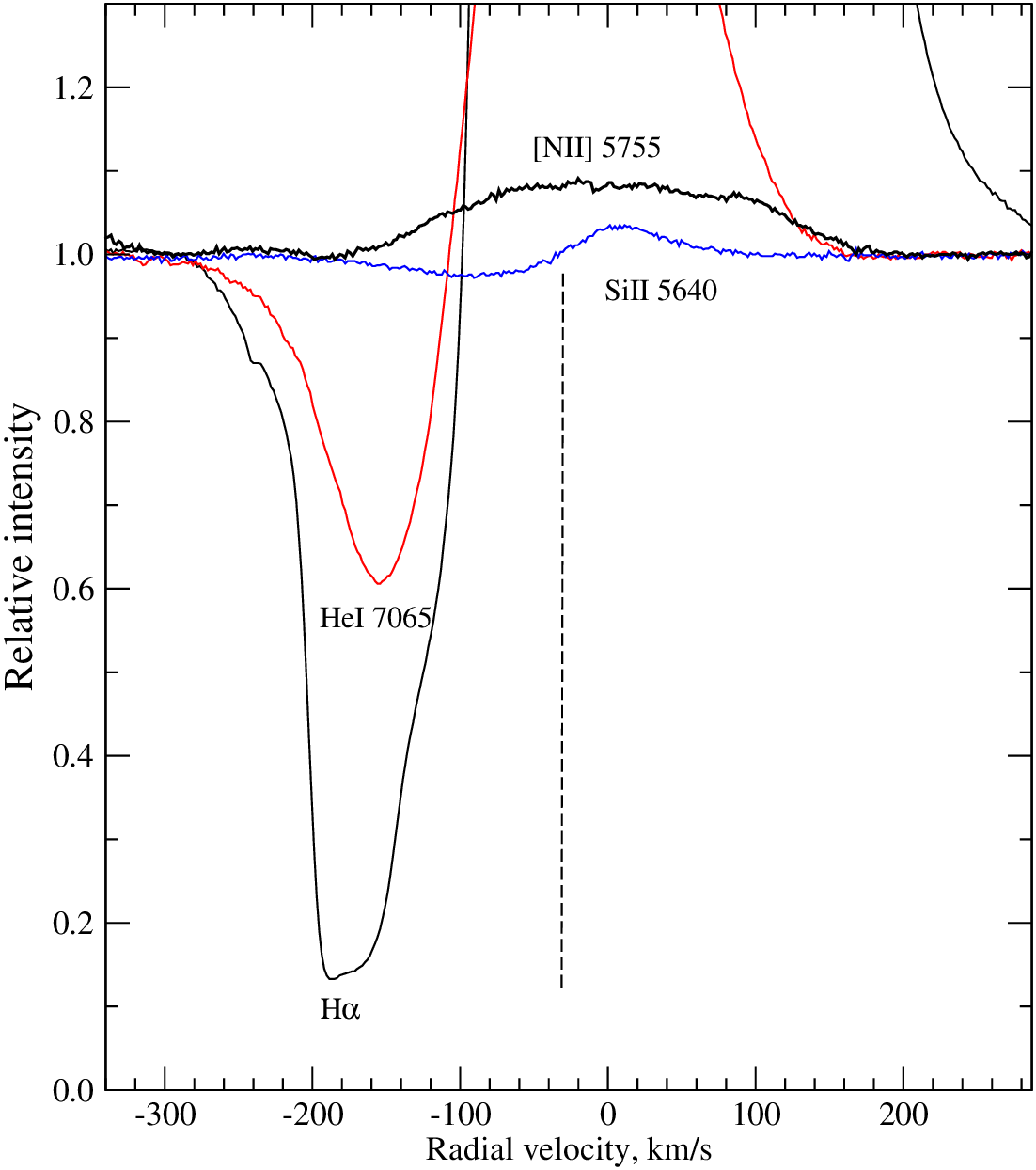}  
\includegraphics[angle=0,width=0.48\textwidth,bb=10 20 550 680,clip]{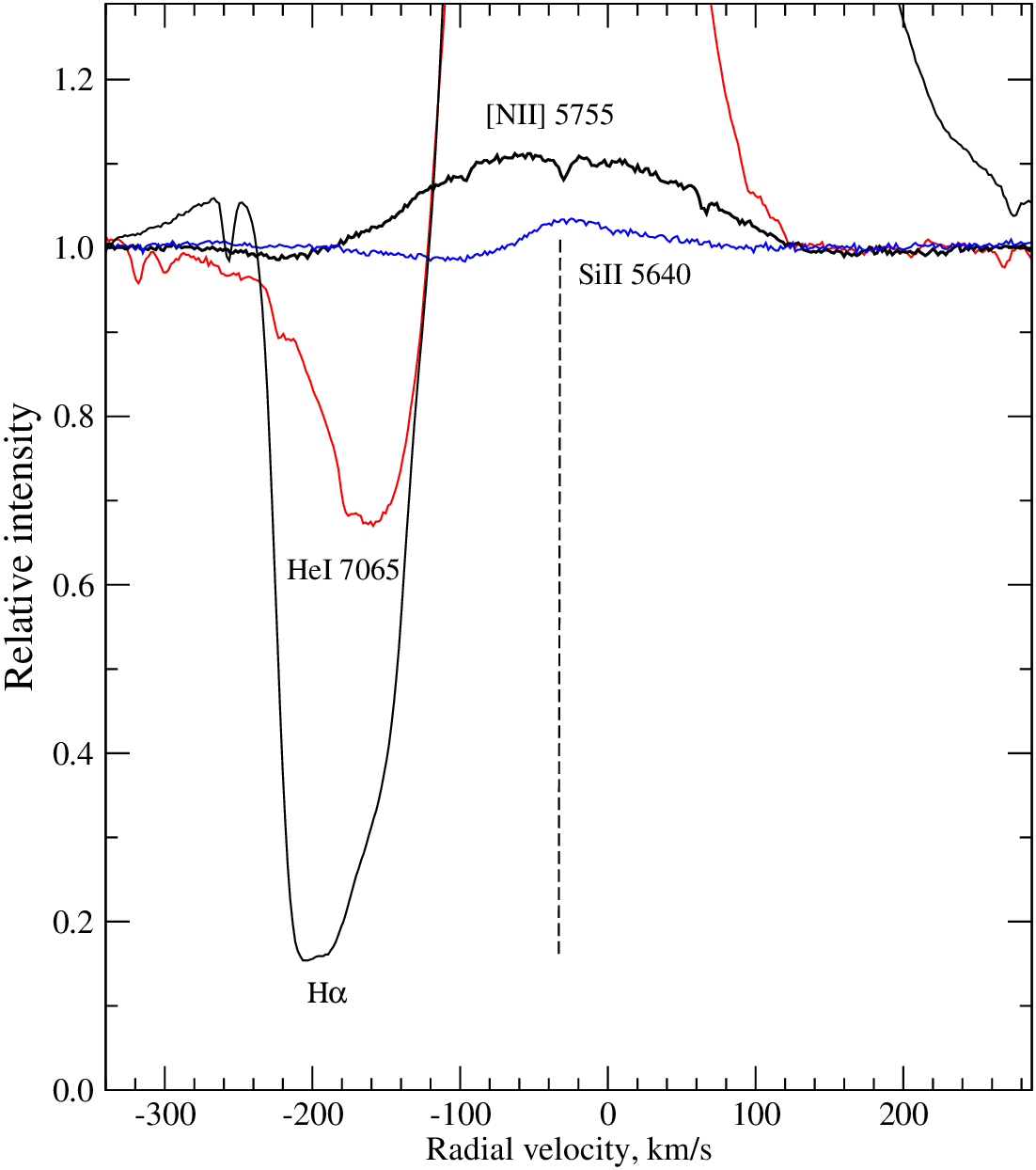}  
\caption{Profiles of selected lines in P\,Cyg spectra obtained on October~26, 2021 (left panel) 
   and September~9, 2022 (right panel): lower part of the H$\alpha$ line -- thin black line, 
   HeI\,7065\,\AA{} -- red line, [N\,II]\,5755\,\AA{} -- bold black line, SiII\,5640\,\AA{} -- blue line.}
\label{Lines}
\end{figure}

{\bf Forbidden emissions}. Figure~7 illustrates the constancy of the position of 
the forbidden emission [N\,II]\,5755\,\AA{}, allowing us to adopt the radial 
velocity value  Vr(5755) as the systemic velocity Vsys=$-34\pm 1.4$\,km/s. 
The authors of~\cite{Smith} determined Vsys=$-38\pm5$\,km/s (LSR$\approx -20\pm 5$\,km/s), 
based on a sample of forbidden emissions in the IR spectrum of P\,Cyg. The
systemic velocity value, LSR$\approx -20$\,km/s, corresponds to the location 
of P\,Cyg in the Perseus arm~\cite{Vallo} and is consistent with the star’s 
distance according to the current value of its parallax.

A close position (Vr$\approx -38$\,km/s) in P\,Cyg spectra is occupied by a 
broad, $\Delta\lambda\approx8$\,\AA{}, emission at a wavelength of 7155\,\AA{} with a
flat top and low intensity, about 3\% above the continuum level. The profile of
this forbidden emission is also shown in Fig.\,7. In the atlas~\cite{Stahl1993},
this detail  is identified as the forbidden emission [FeII]\,7155\,\AA{}, and 
in the spectral  
atlas of MWC\,314 and V1302\,Aql~\cite{Chen1999}, the refined wavelength of the 
emission [FeII]\,7155.14\,\AA{} is given. It is noteworthy that in the spectra
of these  two peculiar hypergiants, MWC\,314 and V1302\,Aql, the relative intensity
 of the  [FeII]\,7155\,\AA{} emission is substantially higher than in the P\,Cyg
spectrum:  I/Icont$\ge1.2$.

For all dates of our observations, we recorded two additional broad emissions with 
flat tops in the P\,Cyg spectrum. One of them at a wavelength of $\approx$7377\,\AA{},
its position corresponds to a velocity of  Vr=$-36.5$\,km/s. In the atlas~\cite{Chen1999}, 
an unidentified feature is also found at this wavelength, I/Icont$\ge1.1$. 
Earlier, such a feature in the P\,Cyg spectrum was identified by the authors~\cite{Stahl1993} 
as the forbidden line [Ni\,II]\,7380\,\AA{}. The peak of the emission profile at 
$\lambda$\,7377\,\AA{} is not entirely flat; the profile is overall similar to
the profiles of forbidden [FeII] lines in the IR spectrum of P\,Cyg, presented
by~\citet{Mizumoto}.  Due to the high S/N ratio, we also  identified a weak, 
I/Icont$\approx 1.05$, emission  of [FeII]\,5261.62\,\AA{}, its position is
Vr=$-31.2$\,km/s  and its width is  $\Delta\lambda\approx8$\,\AA{}.

It should be emphasized that the profile shapes of [FeII]\,7155 and [NiII]\,7377\,\AA{}
emissions significantly differ from the profile of [NII]\,5755\,\AA{} and remind us
of numerous emissions with ``rectangular'' profiles in the spectrum of B[e] star CI\,Cam 
(see, e.g.,~\cite{CICam}). The distinguished positions of these four forbidden emissions
allow us to speak about the stratification of the circumstellar environment, 
due to the presence of low-density structure,  detectable  both kinematically and 
in terms of physical conditions. 

We identified a total of 8 forbidden emissions in the P\,Cyg spectra, and the averaged  
Vr values (and the error of the average) corresponding to their positions in the spectra 
for our nights is indicated in the 3rd row of Table~1. From this data, it follows that 
forbidden emissions,  on average, are formed above the Vsys  level. The
mechanisms of excitation of forbidden emissions (fluorescent excitation due to UV 
radiation from the star itself and impact collisions) in nebulae, including the
circumstellar environment of P\,Cyg, are discussed in~\cite{Smith, Lucy}.

{\bf Photospheric absorptions}. A crucial task in studying spectra of high-luminosity stars, 
filled with circumstellar emissions, is the search and  identification of absorptions
formed in the star’s atmosphere (or in the pseudophotosphere). In the spectrum of P\,Cyg, 
we identified only a small set of unblended absorptions of several ions: Figure~1 contains 
two absorptions, Si\,III\,4819.71 and 4828.95\,\AA{}, from this sample. All identified
absorptions listed in Table~2 are very faint, with depths less than 4--5\,
continuum level. Moreover, some lines (marked in the list with a colon) may be blends. 
The line OII\,6721.388\,\AA{} is free from blending, its depth is $\approx$0.025, and
its corresponding velocity is close to the Vr values from other absorptions in the list. 
Overall, the data in Tables~1 and 2 indicate a stable velocity value for the identified  
photospheric absorptions. The averaged velocity value, Vr(abs)=$-73.8$\,km/s, shifted 
relative to Vsys by $-40$\,km/s, suggests that these absorptions are formed in the
pseudophotosphere. The stability of Vr(abs) allows for a preliminary conclusion about 
the absence of a stellar companion in the P\,Cyg system.

\begin{table*}[ht!]
\medskip
\caption{List of photospheric absorptions identified in the  P\,Cyg spectra and used 
in calculating the average velocity for each date in Table\,1.  Colons mark lines that 
may be blends.}
\begin{tabular}{l| c|  c | c }
\hline
  & \multicolumn{3}{c}{Vr, km/s} \\
\cline{2-4}
Absorbtions& 26.10.2021 &  08.09.2022  &  09.09.2022  \\
\cline{1-4}
  4699.218\,OII   &  -75.66 &          &   -68.38  \\ [-4pt]
  4705.346\,OII   &  -73.27 &   -82.46 &   -73.27  \\ [-4pt]
  4819.712\,SiIII &  -73.31 &   -66.68 &   -72.80  \\ [-4pt]
  4828.951\,SiIII &  -73.42 &   -70.04 &   -67.65  \\ [-4pt]
  4943.005\,OII:  &  -85.20 &   -82.90 &   -77.78  \\ [-4pt]
  5002.703\,NII   &  -67.02 &   -61.24 &   -61.50  \\ [-4pt]
  5010.621\,NII:  &  -84.85 &   -82.83 &   -81.95  \\  [-4pt]
  5032.120\,OII   &  -77.21 &   -79.81 &   -80.05  \\  [-4pt]
  5044.900\,CII:  &  -75.42 &   -76.20 &   -75.01  \\  [-4pt]
  5133.282\,CII   &  -72.61 &   -68.25 &   -76.72  \\  [-4pt]
  5145.165\,CII   &  -77.85 &   -77.75 &   -77.32  \\  [-4pt]
  5452.000\,NII:  &  -70.23 &   -64.76 &   -65.51  \\  [-4pt]
  6721.388\,OII   &  -75.29 &   -74.93 &   -70.96  \\
 \hline
 \end{tabular}
 \end{table*}

{\bf Interstellar features}. The spectrum of P\,Cyg contains numerous absorptions 
formed in the interstellar medium: components of Na\,I D-lines, interstellar
absorption of KI\,7697\,\AA{}, and several Diffuse Interstellar Bands (DIBs).
To illustrate, Figure\,9 shows the profile of the NaI\,5890\,\AA{} line.
This multi-component profile includes: interstellar line ``1'', short-wavelength
absorptions ``2'' and ``3'', and a broad emission ``4''.
The position of the interstellar absorption ``1'' agrees with the radial velocity 
from the positions of interstellar DIBs (see the bottom row of Table 1). The radial
velocity value Vr(KI)=$-11.5$\,km/s from measurements of the position of the 
interstellar line KI\,7697\,\AA{}, shown in Fig.\,9 with a short vertical line,
also agrees with the velocity of the interstellar component ``1''. Weak absorption 
``2'', almost unshifted relative to Vsys, is evidently formed in the extended gas
envelope of P\,Cyg. 
The most interesting components of the profile are the redshifted emission 
``4'' and the broad absorption  ``3'', the velocity range of which from $-85$ to
 $-230$\,km/s allows us to consider the pair of features ``3--4'' as a complex 
P\,Cyg-type profile of the NaI~5890\,\AA{} line. The presence of a P\,Cyg-type
profile in the NaI~D-lines in the spectrum of this star was previously identified~\cite{Markova2000}.
However, with higher-quality spectra, we identified the regions of formation for all
components ``1--4''.

\begin{figure}[ht!]
\includegraphics[angle=0,width=0.8\textwidth,bb=40 50 750 550,clip]{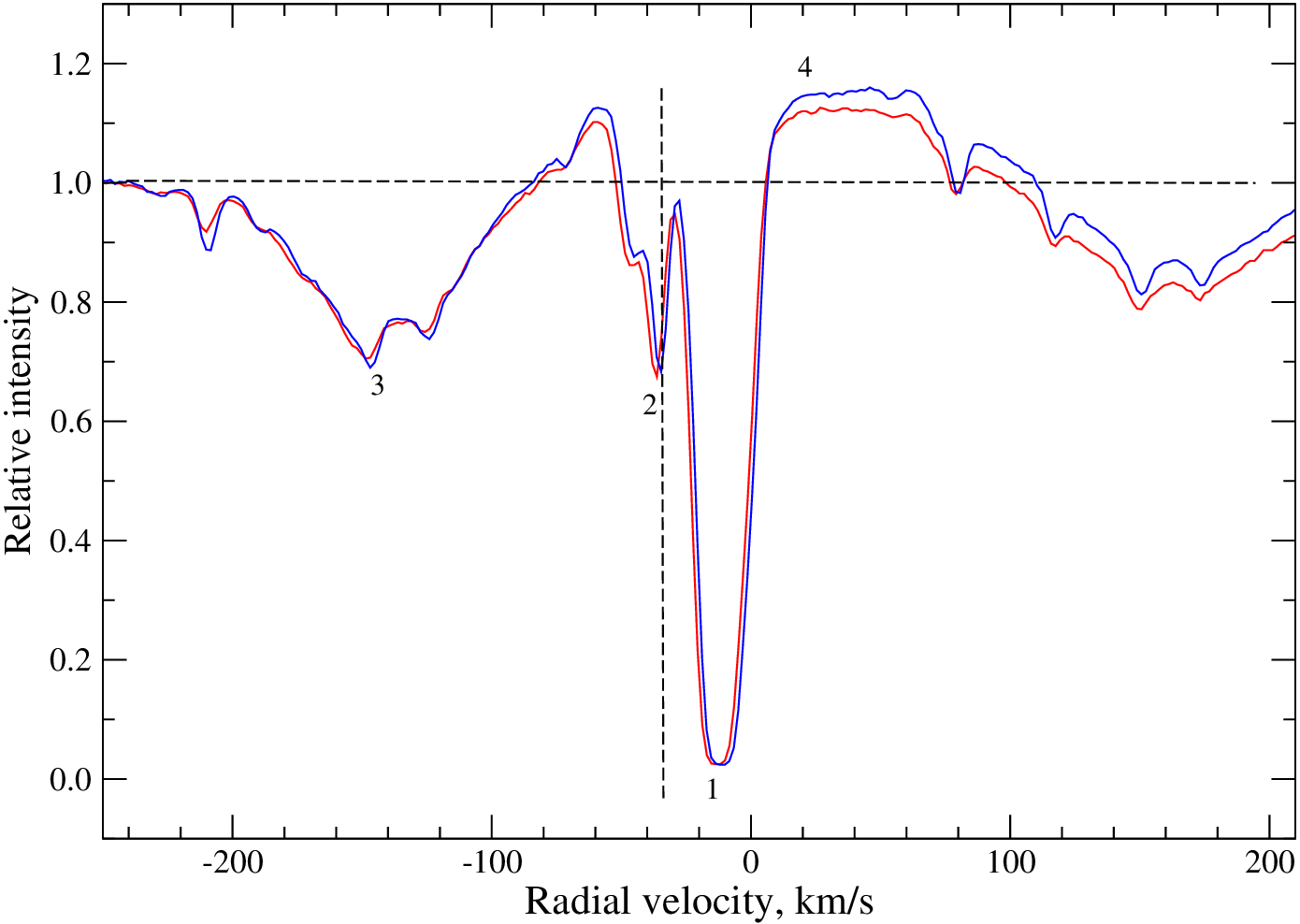}  
\caption{Multi-component profile of the NaI D-line at 5890\,\AA{} in P\,Cyg spectra in 2021 
  and 2022 (red and blue lines, respectively).}
\label{Na5890}
\end{figure}

A similar anomaly of NaI D-line profiles was previously recorded~\cite{V2324Cyg}  
in the spectra of the F-supergiant  V2324\,Cyg (a central star of the IR source 
IRAS\,20572+4919) with an  unclear evolutionary status. The H$\alpha$ line in the 
spectrum of this star has a time-variable P\,Cyg-type profile. Both NaI D-lines 
have P\,Cyg-type profiles and contain deep interstellar absorptions 
(see Fig.\,2 in~\cite{V2324Cyg}). Moreover, in the spectra of V2324\,Cyg, whose 
Galactic  coordinates are close to those of P\,Cyg, the position of NaI 
interstellar absorptions and DIBs,  Vr=$-(12\div$13)\,km/s, agrees  with the 
position of these interstellar features in the spectra of P\,Cyg. 
It is also noted that in the spectrum of V2324\,Cyg, both NaI D-lines have 
a  significantly shifted (by $\Delta$Vr=$−(140\div$225)\,km/s for different 
observation times of this star) wind absorption. Clearly, such a fast wind is 
incompatible with the belonging of V2324\,Cyg  to low-mass post-AGB supergiants.

\begin{table*}[ht!]
\medskip
\caption{Equivalent widths of DIBs and corresponding color excess using calibrations from~\cite{Kos} }
\begin{tabular}{l| c | c }
\hline
$\lambda$\,\AA{} & W$_{\lambda}$,m\AA{} & $E(B-V)$ \\
\cline{1-3}
4963.850   &6   & 0.25  \\ [-5pt]
5780.632   &179 & 0.42  \\ [-5pt]
5797.274   &58  & 0.3   \\ [-5pt]
5849.869   &10  & 0.2   \\ [-5pt]
6196.063   &22  & 0.3   \\ [-5pt]
6203.192   &60  & 0.5   \\ [-5pt]
6269.884   & 24 & 0.3   \\ [-5pt]
6284.198   &364 & 0.4   \\ [-5pt]
6613.766   &44  & 0.25  \\ [-5pt]
6660.815   &19  & 0.4   \\
\hline
 \end{tabular}
 \end{table*}

By measuring the equivalent widths of DIBs listed in Table~3 and applying 
the calibration dependences of $W_{\lambda}$ versus $E(B-V)$ from~\cite{Kos}, we
obtained an average color excess of $E(B-V)=0.34\pm0.03$\,mag for 10  interstellar 
bands. Using the standard ratio R=3.2, we estimate the  interstellar absorption 
to be Av=1.09\,mag.

\section{DISCUSSION OF THE RESULTS}

P\,Cyg is an unique LBV star due to its giant outburst in 1600, which 
resulted in an inflated envelope  -- a rare and historic event. In addition, 
its proximity to the Sun allows for detailed, high-quality observations
across various wavelength regions, providing crucial insights into the 
physics and evolution of massive stars. 

The optical spectrum of P\,Cyg, besides the mentioned features, is notably 
rich in nitrogen features. This includes numerous lines of N\,II ions with 
P\,Cyg-type  profiles, several N\,II photospheric absorptions, and
forbidden emission [N\,II] (see Figs.\,6 and 7). The nitrogen enrichment 
in the spectrum of this evolved  massive star is naturally explained by 
the accumulation of nitrogen in preceding stages of massive star
evolution and subsequent ejection of freshly synthesized chemical elements 
into the circumstellar environment.
Excess nitrogen content has also been found in the atmosphere of the above 
mentioned evolved  massive star V1302\,Aql. As demonstrated by the 
authors~\cite{V1302Aql}, the equivalent widths of N\,II absorptions in the 
spectrum of V1302\,Aql are much higher than for the same lines in the spectrum 
of  the supergiant HD\,13476 with similar parameters. Analogous saturation by 
nitrogen features is revealed by~\citet{Schulte12} in the optical spectrum of 
the LBV candidate  Schulte~12 in  the Cyg~OB2 association.

Several identified details in the spectrum of P\,Cyg allow for a comparison 
with the star MWC\,314, which, due to its extremely high luminosity, was 
considered a LBV candidate~\cite{Mir1998}. However, subsequent studies of 
its high  resolution spectra revealed variability in radial velocity and 
emission profile shapes~\cite{Frasca}. These results led the authors~\cite{Frasca} 
to classify MWC\,314 as a  binary system, including a supergiant with a B[e] 
phenomenon. The results of this study confirm the particular significance 
of high resolution spectroscopy in fixation  the evolutionary status of 
evolved massive stars.  Moreover, the solution to this task is complicated 
by the spectral mimicry of supergiants, as detailed by the authors~\cite{Mimicry}.

The set of spectral features of different natures in the spectrum of P\,Cyg 
provides a considerable diversity of observed profiles. This diversity is 
illustrated in the figures presented in the text, particularly Figs.\,6
and 7, as well as Fig.\,8, which compares profiles of selected lines in 
the spectra of P\,Cyg obtained on October~26, 2021 (left panel) and 
September~9, 2022  (right panel): the lower part of the H$\alpha$ line 
(thin black curve), HeI\,7065\,\AA{} (red curve), [N\,II]\,5755\,\AA{} (bold
black curve), SiII\,5640\,\AA{} (blue curve).

Clearly, the high luminosity near the Humphreys-Davidson limit (see the data 
in the review~\cite{Weis}, specific behaviors of P\,Cyg’s photometric parameters 
over a  time span of over 400~years (see Fig.\,4 in~\cite{Lamers}), and the 
richness of  emissions of various natures in its optical spectrum, including 
forbidden and  permitted lines, serve as indisputable grounds for classifying 
P\,Cyg as an LBV star currently in a dormant state.

\section{Coclusions}

As evident from the publications listed in the Introduction, it is challenging 
to expect the detection of spectral variability in the P\,Cyg spectrum with only
three nights of observations. Nevertheless, we have managed to derive several 
new conclusions. The key outcomes of our study include the following:
\begin{itemize}
\item{} 
 due to the high spectral resolution in P\,Cyg’s spectra, the majority of 
spectral details of various types have been identified, encompassing photospheric
absorptions of CNO triad ions, pure metal emissions, forbidden emissions 
(particularly [N\,II], [S\,II], [NiII]), and lines with P\,Cyg-type profiles 
featuring wind absorption positions in a broad range of radial velocities,
$\Delta{\rm Vr}=-(140\div250)$\,km/s.
\item{}  the systemic velocity is fixed at  Vsys=$-34\pm1.4$\,km/s based on 
the stable positions of four forbidden emissions, including [N\,II]\,5755\,\AA{}.
\item{} no forbidden emissions, such as [O\,I]\,5577,
   6300, 6363 Å, [OIII]\,4959 and 5007\,\AA{}, as well as high excitation
   HeII\,4686\,\AA{} emission, were detected.
\item{} the stability of the position of revealed photospheric absorptions, with 
an  average velocity,  Vr(abs)=$-73.8$\,km/s, lower than Vsys by $-40$\,km/s
suggests that these absorptions form in the pseudophotosphere. This result, 
indicating  the absence of a companion in the P\,Cyg system, requires further 
observations for confirmation.
\item{} DACs were observed only in the profiles of H\,I and He\,I lines.
\item{} variability of the intensity of peak emission values, V/R ratios, 
  and their positions was observed for all features with P\,Cyg-type profiles.
\item{} the regions of formation for all four components of Na I D-line profiles 
    are identified.
\item{} interstellar absorption, Av=1.09\,mag, was determined based on the 
   intensities of a set of DIBs.
\end{itemize}

The results obtained in the initial stage of spectral monitoring of P\,Cyg 
allow us to assert the effectiveness of our approach in addressing the 
task of searching and studying the variability of the star’s complex spectrum.
The extremely rare reocurrence of giant outbursts indicates the need for 
prolonged spectral monitoring with high spectral resolution across a broad 
wavelength range.

\section*{ACKNOWLEDGMENTS}
This study used information from astronomical databases SIMBAD, VALD, 
SAO/NASA ADS, and Gaia~DR3.

\section*{FUNDING}

Observations on the 6-m telescope of the Special Astrophysical
Observatory of the Russian Academy of Sciences were supported by the Ministry 
of Science and Higher Education of the Russian Federation.

\end{document}